\documentclass[12pt,preprint]{aastex}
\begin{document}

\title{CCD photometric study of the W~UMa-type binary II CMa
in the field of Berkeley~33 }

\author{Liu L.\altaffilmark{1,2,3} and Qian S.-B.
\altaffilmark{1,2};\\ Zhu, L.-Y.\altaffilmark{1,2,3},
 He, J.-J.\altaffilmark{1,2,3}, Yuan, J.-Z.\altaffilmark{1,2,3}, Dai, Z.-B.\altaffilmark{1,2,3}, Liao, W.-P.\altaffilmark{1,2,3}, Zhang, J.\altaffilmark{1,2,3}}

\altaffiltext{1}{National Astronomical Observatories/Yunnan
Observatory, Chinese Academy of Sciences, \\
P.O. Box 110, 650011 Kunming, P.R. China\\
 LiuL@ynao.ac.cn\\
qsb@netease.com}

\altaffiltext{2}{United Laboratory of Optical Astronomy, Chinese
Academy of Sciences (ULOAC),\\ 100012 Beijing, P. R. China}

\altaffiltext{3}{Graduate School of the Chinese Academy of
Sciences,\\ 100012 Beijing, P. R. China }


\begin{abstract}

The CCD photometric data of the EW-type binary, II~CMa, which is a
contact star in the field of the middle-aged open cluster
Berkeley~33, are presented. The complete R light curve was obtained.
In the present paper, using the five CCD epochs of light minimum (
three of them are calculated from Mazur et al. (1993)'s data and two
from our new data), the orbital period P was revised to 0.22919704
days. The complete R light curve was analyzed by using the 2003
version of W-D (Wilson-Devinney) program. It is found that this is a
contact system with a mass ratio $q=0.9$ and a contact factor
$f=4.1\%$. The high mass ratio ($q=0.9$) and the low contact factor
($f=4.1\%$) indicate that the system just evolved into the marginal
contact stage.
\end{abstract}

\section{Introduction}

II~CMa, which was first discovered as a byproduct by Mazur et al.
(1993) when they searched for variable stars in the intermediate age
open cluster Berkeley~33, is a W~UMa-type contact binary in the
field of the cluster (Berkeley~33,
$\alpha_{2000.0}=06^{h}57^{m}42^{s}$,
$\delta_{2000.0}=13^\circ13'00''$, either named as Ruprecht~7). They
marked it as $\texttt{v}_1$, and classified it into EW-type. It was
named as II~CMa by the latest version of the General Catalogue of
Variable Stars (GCVS for short, Samus et al. 2004). It is very clear
to see a O'Connell effect (O'Connell, 1951) in Mazur et al.'s V, I
lights. The amplitude in V \& I magnitude were both 0.5 mag
according to Mazur et al. 1993. They also gave its period as 0.2292
days, (V-I) as 1.44, (B-V) as 1.3. They concluded that II~CMa is
probably a foreground contact binary, of later type, similar to the
well-studied system CC~Com and V523~Cas. The other research about it
came from Rucinski (1998). Having used $E(B-V)_{Be33}=0.7$,
$(B-V)_{II~CMa}=1.30$, Rucinski (1998) pointed out the contact
binary star II~CMa is definitely not members of the cluster Be~33.

\section{Observations for II~CMa}
After observed on two nights (February 19, 21, 2007) with the PI1024
TKB CCD photometric system attached to the 1.0-m reflecting
telescope at the Yunnan Observatory in China, the data in R band of
II~CMa were obtained. The R color systems used are close to the
standard UBVRI system and the effective field of view of the
photometric system is $6.5\times6.5$ arc min at the telescope's
Cassegrain focus. The integration time for each image is 150\,s.
PHOT (measure magnitudes for a list of stars) of the aperture
photometry package of IRAF was used to reduce the observed images.

The observations in the R passband completely cover the orbital
phase. We calculated the phase of the observations with the linear
ephemeris (Eq.1) given in next section, and plotted the R light in
Figure 1. Our original data in the R band are listed in Table 1. The
light curve is continuous and possesses a very small difference
between the depths of minima which reveal the system was tidally
distorted and both components' temperature may be similar. The
maximum brightness is 15.22mag in V passband and the amplitude of
the light variation is $\sim0.45$\,mag. Since the two minima in the
curve show a sharp variation, those properties suggest that the
system has a low orbital inclination and the size of the two
components might be similar. We also plotted the comparison and the
check star's corresponding curve in Figure 1. It is a straight line
showing that the comparison star is constant and the variation of
the curve upon plate in Figure 1 truly came from II~CMa.
Corresponding coordinates of the binary, comparison and check star
are listed in Table 2. We chose the comparison star as close to the
variable as it can where the range of air-mass difference between
both stars was very small ($\sim0.0007$). Therefore, extinction
correction was not made. It is shown in this figure that the data
are high quality except for several data points and the light
variation is typical of EW-type. Due to the light minimum is
symmetric, a parabolic fitting was adopted to determine the times of
minimum light with a least square method. Two epochs of light
minimum were obtained in all and are listed in the last two rows of
Table 3.

There are some obvious variations in the light curves, such as
O'Connell effect, eclipse depth at minimum. It is very clear to see
a O'Connell effect (O'Connell, 1951) in Mazur et al.'s V, I lights.
Namely the magnitude of 0.75 phase is about 0.022 mag bright than
that of the 0.25 phase in I light. In V light, the O'Connell effect
is the same as I light, but the value is slightly smaller than 0.022
mag, it is about 0.02 mag. We can call that as negative O'Connell
effect. However, it disappeared now. Another variation in the light
curve is that the amplitude in V magnitude was 0.5 mag according to
Mazur et al. 1993, while that in R is 0.45 mag now. Although
observed in different passband, noting that wavelength of R band is
between that of V band and I band, we have sufficient reasons to
believe the light curve changes. On the other hand, this kind of
changes are frequently seen in W~UMa-type binaries. For instance,
FG~Hya (Qian \& Yang, 2005), AH~Cnc (Qian et al. 2006), EQ~Tau (Yang
\& Liu, 2004), CU~Tau (Qian et al. 2005), AD~Cnc (Yang \& Liu,
2002a), QX~And (Qian et al. 2007), CE~Leo (Yang \& Liu, 2002b, Kang
et al. 2004), BX~Peg (Lee et al. 2004).

\begin{table*}
\centering \caption{Photometric Data in the R band for II~CMa
observed with the 1.0 meter telescope at Yunnan observatory on
February 19, 21, 2007}
\begin{tabular}{llllllllll}\hline\hline
JD.Hel.     &${\Delta}m$& JD.Hel.   &${\Delta}m$ &JD.Hel.    &${\Delta}m$&JD.Hel.  &${\Delta}m$&JD.Hel.  &${\Delta}m$\\
2454100+    &         &  2454100+   &         &    2454100+  &        &2454100+    &         &2454100+    &          \\
\hline
51.0623 & 1.374  &  51.1160 & 1.392  & 51.1686 & 1.480   &  51.2226 & 1.382  &  53.0843 & 1.618  \\
51.0646 & 1.373  &  51.1181 & 1.409  & 51.1707 & 1.476   &  51.2249 & 1.388  &  53.0863 & 1.656  \\
51.0673 & 1.351  &  51.1203 & 1.406  & 51.1728 & 1.458   &  51.2269 & 1.379  &  53.0882 & 1.657  \\
51.0693 & 1.348  &  51.1224 & 1.430  & 51.1749 & 1.440   &  51.2290 & 1.406  &  53.0903 & 1.695  \\
51.0714 & 1.347  &  51.1245 & 1.464  & 51.1770 & 1.466   &  51.2312 & 1.398  &  53.0922 & 1.708  \\
51.0735 & 1.326  &  51.1266 & 1.494  & 51.1791 & 1.437   &  51.2333 & 1.403  &  53.0942 & 1.735  \\
51.0756 & 1.330  &  51.1286 & 1.510  & 51.1812 & 1.438   &  51.2354 & 1.437  &  53.0962 & 1.729  \\
51.0777 & 1.336  &  51.1307 & 1.525  & 51.1837 & 1.401   &  51.2376 & 1.461  &  53.0981 & 1.744  \\
51.0798 & 1.333  &  51.1328 & 1.556  & 51.1858 & 1.383   &  51.2398 & 1.494  &  53.1001 & 1.716  \\
51.0818 & 1.326  &  51.1349 & 1.580  & 51.1879 & 1.362   &  53.0523 & 1.324  &  53.1021 & 1.686  \\
51.0839 & 1.325  &  51.1370 & 1.585  & 51.1900 & 1.349   &  53.0544 & 1.348  &  53.1041 & 1.679  \\
51.0860 & 1.313  &  51.1391 & 1.638  & 51.1921 & 1.338   &  53.0564 & 1.336  &  53.1060 & 1.634  \\
51.0881 & 1.318  &  51.1412 & 1.673  & 51.1942 & 1.317   &  53.0584 & 1.371  &  53.1080 & 1.619  \\
51.0902 & 1.326  &  51.1433 & 1.703  & 51.1962 & 1.320   &  53.0603 & 1.359  &  53.1100 & 1.589  \\
51.0922 & 1.307  &  51.1455 & 1.720  & 51.1983 & 1.314   &  53.0623 & 1.391  &  53.1122 & 1.576  \\
51.0944 & 1.311  &  51.1476 & 1.691  & 51.2004 & 1.312   &  53.0643 & 1.381  &  53.1143 & 1.548  \\
51.0964 & 1.316  &  51.1497 & 1.686  & 51.2033 & 1.320   &  53.0664 & 1.414  &  53.1164 & 1.521  \\
51.0985 & 1.332  &  51.1518 & 1.666  & 51.2054 & 1.313   &  53.0684 & 1.414  &  53.1185 & 1.496  \\
51.1006 & 1.326  &  51.1539 & 1.665  & 51.2075 & 1.309   &  53.0703 & 1.441  &  53.1205 & 1.485  \\
51.1027 & 1.330  &  51.1560 & 1.635  & 51.2096 & 1.312   &  53.0723 & 1.443  &  53.1225 & 1.481  \\
51.1055 & 1.350  &  51.1581 & 1.620  & 51.2117 & 1.339   &  53.0744 & 1.473  &  53.1244 & 1.456  \\
51.1076 & 1.316  &  51.1601 & 1.614  & 51.2140 & 1.308   &  53.0763 & 1.487  &  53.1264 & 1.429  \\
51.1098 & 1.332  &  51.1622 & 1.565  & 51.2161 & 1.328   &  53.0784 & 1.518  &  53.1284 & 1.424  \\
51.1118 & 1.334  &  51.1644 & 1.538  & 51.2182 & 1.347   &  53.0803 & 1.554  &          &        \\
51.1139 & 1.355  &  51.1665 & 1.514  & 51.2205 & 1.354   &  53.0823 & 1.559  &          &        \\
\hline\hline
\end{tabular}
\end{table*}

\begin{figure}
\begin{center}
\includegraphics[angle=0,scale=1 ]{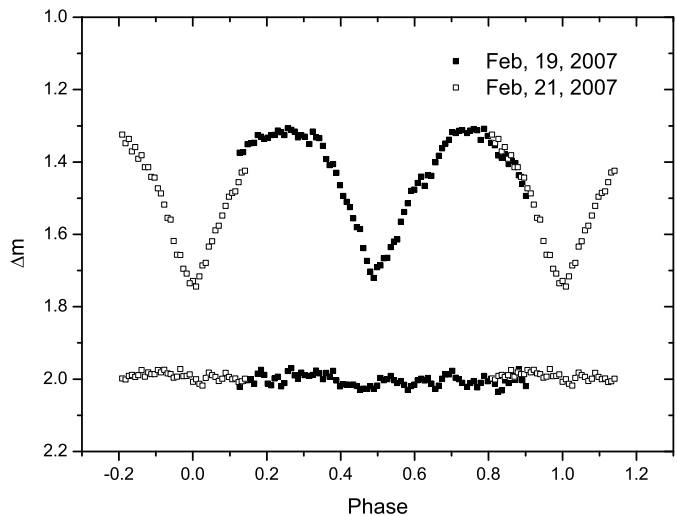}\caption{CCD data in the
R band of II~CMa observed at 19 and 21 February, 2007.} \end{center}
\end{figure}

\begin{table}
\centering\caption{Coordinates of II~CMa, the comparison, and the
check stars.}
\begin{tabular}{ll}\hline\hline
Stars & 2MASS \\\hline
II~CMa   & 06573640-1314364\\
The comparison  & 06573261-1314336\\
The check       & 06573952-1312477\\
\hline\hline
\end{tabular}
\end{table}

\section{New ephemeris of II CMa}

There were too lack of studies about II~CMa to investigate its
period changes. Consequently, we can only revise the linear
ephemeris. By setting the initial period value as 0.2292 days
declared by Mazur et al. (1993) and combining the three times of
minimum calculated from Mazur et al. (1993)'s data and two of ours,
we determined a new linear ephemeris as follows,
\begin{eqnarray}
{\rm Min.~I}&=&2454153.0961(\pm0.0008)\nonumber\\
    & &+0.^{d}22919704(\pm0.00000004)\times{E},
\end{eqnarray}
and it is suggested that the much more accurate period should be
0.22919704 days. The correction about $-0.00000296$ days were made.
After amended, that new linear ephemeris given by us is more precise
than before, which can forecast the times of minimum about II~CMa in
the future. The phases of the observations and the $(O-C)$ values
were calculated through that Equation (See Table 3).

\begin{table*}
\begin{center}\caption{CCD times of light minimum for II~CMa.}
\begin{tabular}{llllcll}\hline\hline
JD.Hel.    & Error (days) & Filters & Min.&  E        & $(O-C)    $  &  Origin\\
\hline
2448253.2205 &$\pm0.0009$   & V       & II  &$-$25741.5 & $+$0.0000    & Mazur et al. (1993)'s data \\
2448286.1117 &$\pm0.0011$   & I       & I   &$-$25598   & $+$0.0014    & Mazur et al. (1993)'s data \\
2448291.1511 &$\pm0.0011$   & I       & I   &$-$25576   & $-$0.0015    & Mazur et al. (1993)'s data \\
2454151.1480 &$\pm0.0004$   & R       & II  &$-$8.5     & $+$0.0000    & The present paper\\
2454153.0961 &$\pm0.0003$   & R       & I   &   0       & $+$0.0000    & The present paper\\
\hline\hline
\end{tabular}
\end{center}
\end{table*}

\section{Photometric Solution}
Neither the photometric nor the spectral parameters of II~CMa have
been derived before. So, the initial values must be estimated, such
as temperatures, mass ratio, and otherwise. Because it appears in
the field of the cluster Be~33, they might got similar extinction.
So we need to know some information about the cluster.

There are several authors had study the middle-aged open cluster
Berkeley~33 (Mazur et al. 1993, Lata et al. 2002, Hasegawa et al.
2004, Carraro et al. 2005). Mazur et al. (1993), Lata et al. (2002)
derived the cluster's $E(B-V)$ and distance modulus were 0.70 and
15.5 respectively. However, recently, Hasegawa et al. (2004),
Carraro et al. (2005) have suggested that the corresponding
parameters mentioned above were 0.3 and 15.0. It is obvious that
there are two groups of measured parameters. The data published in
earlier years are group one, and in later years are definitely group
two. In order to compare these two different groups, we listed them
in Table 4.

However, the distance modulus of Hasegawa et al. (2004) is 14.2, and
the age is 1.3Gyr, which are very different from the other three
values. As they said, that was the most difficult cluster in their
sample to fix the cluster parameters, so that their efforts might
not give a convincing fit.
\begin{table*}
\centering\caption{The different parameters of Berkeley~33.}
\begin{tabular}{llllll}\hline\hline
 $E(B-V)$ & $E(V-I)$ & $(m-M)$ & Age(Gyr) &$[Fe/H]$&Reference\\
 \hline
  0.70&    & 15.5& 0.7 &-0.6 &Mazur et al. 1993 \\
  0.70&    & 15.6& 0.7 &     &Lata et al. 2002 \\
  0.30&0.50& 14.2& 1.3 &0.008&Hasegawa et al. 2004 \\
  0.30&0.47& 15.0& 0.8 &0.019&Carraro et al. 2005 \\
\hline\hline
\end{tabular}
\end{table*}

Considering the two different group parameters (Table 4), the
discussions were under two cases, namely Case A and Case B.

 (i)  Case A, the color excess is $E(B-V)=0.7$, and the corresponding
temperature is $5800K$ by used the well known experiential formula,
\begin{eqnarray}
T = 8540/[(B-V)_0 + 0.865],
\end{eqnarray}
where $(B-V)_0 = (B-V)-E(B-V)$.

 (ii)  Case B, the color excess is $E(B-V)=0.3$, and the corresponding
temperature is $4579K$.

On the other hand, as Eggen having pointed out in 1967 (Eggen,
1967), for contact binaries, there is a strict relationship between
color and period, namely $C=-0.50-2.26\rm log P$. Hence, we can
obtain the color-index as 0.946, which much better corresponds to
the Case B. As a result of that, we assumed the primary component's
temperature is $4579K$.

In order to get a probable value of the mass ratio $q$, a q-search
method with the 2003 version of the W-D program (Wilson \& Devinney,
1971; Wilson, 1990, 1994; Wilson \& Van Hamme, 2003) was used
(Figure 2). Firstly, we fixed q to 0.2, 0.3, 0.4 and so on in both
of Case A and Case B, as figure 4 shows. It can be seen that the
value of $q$ is a range from 0.3 to 0.8. Thought about the
characters of the light curve, as mentioned above (Section 2), we
incline to the value $q=0.9$.

During the solution, the bolometric albedo $A_1=A_2=0.5$ (Rucinski
1969) and the values of the gravity-darkening coefficient
$g_1=g_2=0.32$ (Lucy 1967) were used, which correspond to the common
convective envelope of both components. Limb-darkening coefficient
of 0.499 in R under Case A and 0.645 in R under Case B were used,
according to Claret \& Gimenez (1990). The adjustable parameters
were, the mass ratio $q$; the orbital inclination $i$; the mean
temperature of star 2, $T_2$; the monochromatic luminosity of star
1, $L_{1R}$; and the dimensionless potential of star 1
($\Omega_1=\Omega_2$, mode 3 for contact configuration). The results
of the photometric solutions are listed in Table 5 and the
theoretical light curves computed with those photometric elements
are plotted in Figure 3. Meanwhile, the geometrical structure of
II~CMa is displayed in Figure 4.

\begin{table*}
\centering\caption{Photometric Solutions for II~CMa.}
\begin{tabular}{lcl}\hline\hline
Parameters              &  Photometric elements &  errors  \\
 \hline
$g_1=g_2$               &    0.32               & assumed  \\
$A_1=A_2$               &    0.50               & assumed \\
$x_{1R}=x_{2R}$         &    0.645               & assumed\\
$T_1$                   &    4579K              & assumed\\
$q$                     &    0.9000             & assumed\\
$\Omega_{in}$           &    3.5856             & --    \\
$\Omega_{out}$          &    3.0880             & --     \\
$T_2$                   &    4475K              & $\pm29$K \\
$i$                     &    68.276             & $\pm0.415$ \\
$L_1/(L_1+L_2)(R)$      &    0.5534             & $\pm0.0405$ \\
$\Omega_1=\Omega_2$     &    3.5653             & $\pm0.0192$\\
$r_1(pole)$             &    0.3676             & $\pm0.0030$\\
$r_1(side)$             &    0.3871             & $\pm0.0039$\\
$r_1(back)$             &    0.4189             & $\pm0.0058$\\
$r_2(pole)$             &    0.3500             & $\pm0.0032$\\
$r_2(side)$             &    0.3676             & $\pm0.0040$\\
$r_2(back)$             &    0.4002             & $\pm0.0062$\\
the contact factor~$f$     &   $4.1\,\%$         & $\pm0.8\,\%$\\
$\sum{\omega_i(O-C)_i^2}$ &   0.006279                        \\
\hline\hline
\end{tabular}
\end{table*}

\begin{figure}
\begin{center}
\includegraphics[angle=0,scale=1 ]{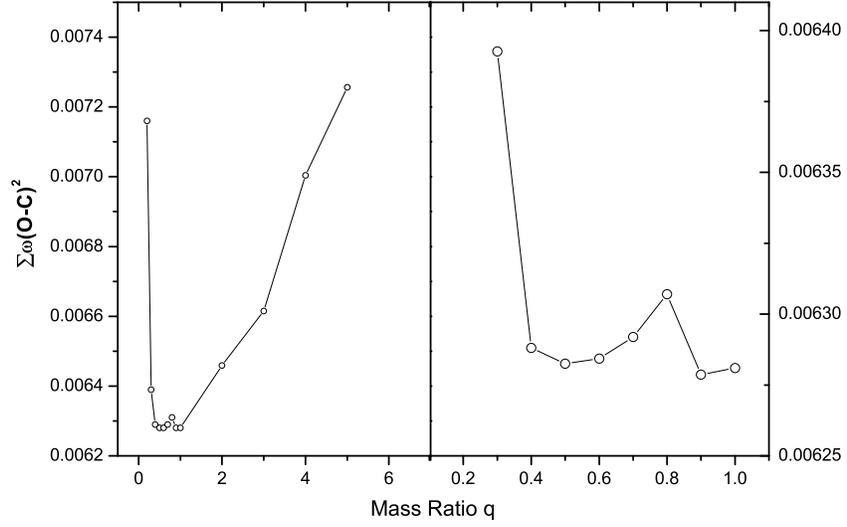}\caption{The relation
between q and $\Sigma$ for II~CMa, the right panel is the megascopic
corresponding field in the left panel.}\end{center}
\end{figure}

\begin{figure}
\begin{center}
\includegraphics[angle=0,scale=1 ]{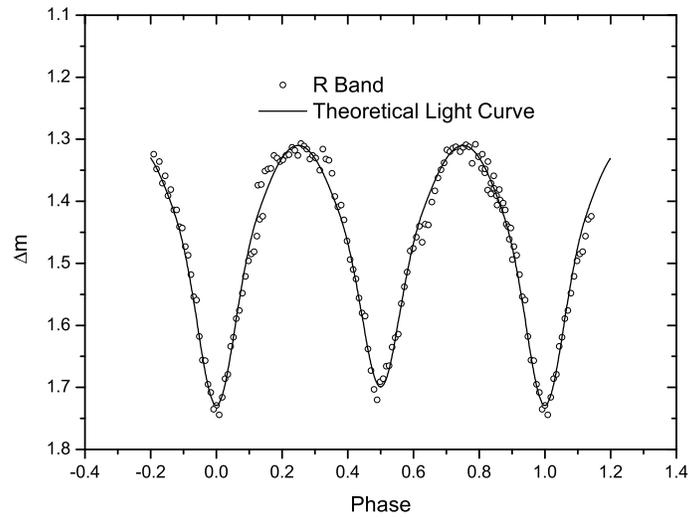}\caption{Observed and
theoretical light curves in the R band for II~CMa.} \end{center}
\end{figure}

\begin{figure}
\begin{center}
\includegraphics[angle=0,scale=1 ]{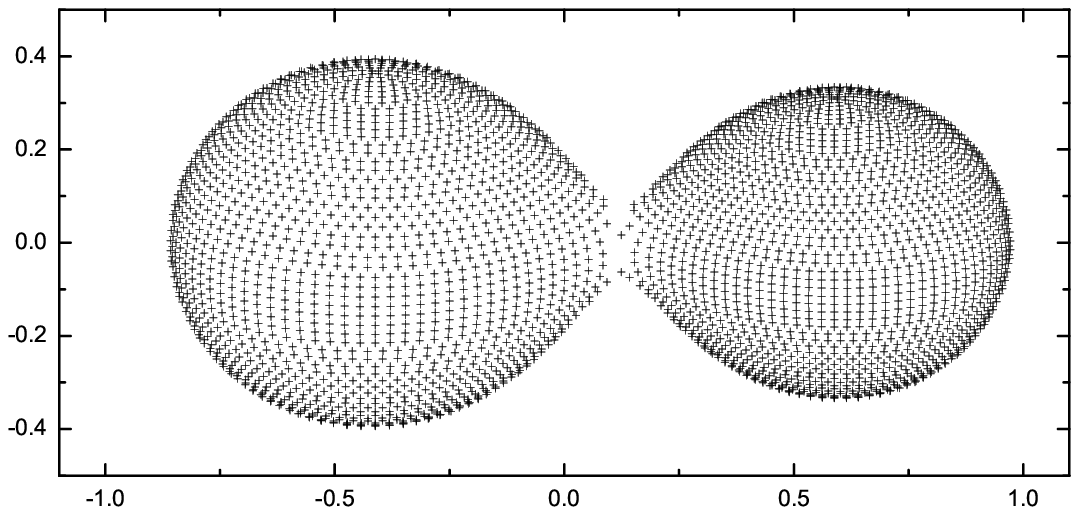}\caption{Geometrical
structure of the high mass ratio shallow contact binary II CMa.}
\end{center}
\end{figure}

\section{Discussion and Conclusion}
After two nights' (Feb 19, 21, 2007) observations, the complete R
light was obtained. It is very easy to find that there exist some
distinct changes about O'Connell effect (O'Connell, 1951), which
consist in many W~UMa-type binaries such as FG~Hya (Qian \& Yang,
2005), AH~Cnc (Qian et al. 2006), EQ~Tau (Yang \& Liu, 2004), CU~Tau
(Qian et al. 2005), AD~Cnc (Yang \& Liu, 2002a), QX~And (Qian et
al., 2007), CE~Leo (Yang \& Liu, 2002b, Kang et al. 2004), BX~Peg
(Lee et al. 2004), and so on. A O'Connell effect (O'Connell, 1951)
of II~CMa was in Mazur et al.'s V, I lights. The magnitude of 0.75
phase is about 0.022 bright than that of the 0.25 phase in I light.
In V light, the O'Connell effect is the same as I light, but the
value is slightly smaller than 0.022, it is 0.02. But we did not
find distinctness in our light curve.

Using the five times of minimum of II~CMa listed in Table 3, the
orbital period was revised as 0.22919704 days. The system is a high
mass ratio shallow contact binary with $q=0.9$, $f=4.1\%$. These
suggest that the system is a marginal contact binary. There are a
hundreds of difference temperatures between the two components, that
phenomena was usually in contact binaries.

High mass ratio, low contact factor contact systems, which are much
more interest and important, are the key to understanding the
evolution status of a close binary from the near contact phase to
the contact phase. Nearly, almost all acceptable models of contact
binary foretell that while contact age is increasing, $q$ is
decreasing. This is on account of ineluctable mass transfer from the
less mass component to the more one when the system came to the
contact stage. Although there are two assumed cases about the
system, both the mass ratio and the low contact factor suggest that
II~CMa is just come into contact so that it has younger contact age.
This kind of contact binaries which were called marginal contact
binaries ($f\leqslant 10\%$) are indicators of evolution time scale
into the contact stage. For example, V803 Aql (Samec et al. 1993),
FG Sct (Bradstreet, 1985), RW PsA (Lucy \& Wilson, 1979), XZ Leo
(Niarchos et al. 1994), S Ant (Russo et al. 1982). These marginal
contact binaries' parameters are list in Table 6.
\begin{table}
\begin{scriptsize} \caption{Parameters of some marginal contact
binaries ($f\leqslant 10\%$)}
\begin{tabular}{llllllll}
\hline\hline
Name     & Period       &$T_1$(K)&$T_2$(K)& $i(^\circ)$   &$q$    &$f\%$  & ref.$^{*}$\\
\hline
V803 Aql & 0.2634       &4600    &4594    & 82.910  &1.0    &8         &(1)  \\
FG Sct   & 0.27057192   &4800    &4662    & 89.9    &1.273  &8.5       &(2)  \\
RW PsA   & 0.36045011   &5600    &5325    & 77.45   &0.813  &7         &(3)  \\
XZ Leo   & 0.48773526   &7850    &7044    & 72.035  &0.726  &7         &(4)  \\
S Ant    & 0.6483358    &7800    &7340    & 69.49   &0.590  &9         &(5)  \\
II CMa   & 0.22919704   &4579    &4475    & 68.276  &0.900  &4.1       &(6)  \\
\hline
\end{tabular}\\{References in Table 6: (1) Samec et al. (1993); (2) Bradstreet (1985); (3)
Lucy \& Wilson (1979); (4) Niarchos et al. (1994); (5) Russo et al.
(1982); (6) The present authors.}
\end{scriptsize}\end{table}

II CMa would not be a member of Be33 for two reasons. Firstly,
according Rucinski's (1998) result, $M_v=-4.44\rm
logP+3.02(B-V)_0+0.12$, its distance modular is 9.32, which
indicates this system is in front of the cluster in space. Secondly,
assumed it is a cluster member, noticing that the binary II~CMa
appears at the red giant Branch in the cluster's Color-Magnitude
Diagram, it should be a red giant system. But based on the theory of
structure and evolution of stars, it hardly to explain how such
system who possesses a typical EW-type light curve be born. So, we
prefer that II CMa is a front field star. It is remarkable that II
CMa and the cluster have same interstellar extinction (see previous
sections), which may imply there is a big extinction matter in front
of them.

Colligating the reasons above, we can say that the W~UMa-Type high
mass ratio shallow contact binary in the intermediate-aged open
cluster Berkeley~33 field II~CMa is a very interesting object to be
study. It will be helpful to understand something about marginal
contact binaries.

 Eventually, since the eclipse binary is partial
and absolute physical dimensions are still uncertain yet, it can not
take the purely light solution as a definition. It is just a
reasonable result. Only will precision radial-velocity curves put
our result on a strong footing. In the future, for identify its
identity strictly, a precise spectra observation is needed, which
may help us to make sure weather it a member or not. Speaking for
itself, this would be the most important element we need to know,
otherwise all the works purely stood on a assumption. And another
thing, we should accumulate the times of its minimum in order to
investigate its orbital period changes, which can open out its
evolution stage. In a word, it is essential to monitor the every
interesting high mass ratio low contact factor contact binary system
II~CMa.

\vskip 0.3in \noindent Acknowledgments: This work was partly
supported by Yunnan Natural Science Foundation (No.2005A0059M) \&
the Chinese Natural Science Foundation(10573032, 10573013, and
10433030). New observations of II~CMa were obtained with the 1.0-m
telescope at Yunnan Observatory. Thanks to the anonymous referee who
given us useful comments and cordial suggestions, which helped us to
improve the paper greatly.

\end{document}